\documentclass[aps,prc,superscriptaddress,showpacs,amsmath,amssym,
nofootinbib,twocolumn, aastex]{revtex4-1}

\usepackage{array}
\usepackage{graphicx}
\usepackage{epstopdf}
\usepackage{color}
\usepackage{bm}
\usepackage{cmap}
\usepackage{nicefrac}
\usepackage{xfrac}
\usepackage{cleveref}
\usepackage{booktabs}
\usepackage{tabularx}

\newcommand{\beq}{\begin{equation}}
\newcommand{\eeq}{\end{equation}}

\newcommand{\slfrac}[2]{\left.#1\middle/#2\right.}

\begin{document}

\title{Effective Potential and Interdiffusion
in Binary Ionic Mixtures}
\author{M.\ V.\ Beznogov}
\affiliation{St.~Petersburg Academic University, 8/3 Khlopina Street,
St.~Petersburg 194021, Russia}
\author{D.\ G.\ Yakovlev}
\affiliation{Ioffe Physical Technical Institute, 26 Politekhnicheskaya,
St.~Petersburg 194021, Russia}

\date{\today}

\begin{abstract}
We calculate interdiffusion coefficients in a two-component, weakly
or strongly coupled ion plasma (gas or liquid, composed of two ion
species immersed into a neutralizing electron background). We use an
effective potential method proposed recently by Baalrud and Daligaut
[PRL, {\bfseries 110}, 235001, (2013)]. It allows us to extend the standard
Chapman-Enskog procedure of calculating the interdiffusion
coefficients to the case of strong Coulomb coupling. We compute
binary diffusion coefficients for several ionic mixtures and fit
them by convenient expressions in terms of the
generalized
Coulomb
logarithm. These fits cover a wide range of plasma parameters
spanning from weak to strong Coulomb couplings. They can be used to
simulate diffusion of ions in ordinary stars as well as in white
dwarfs and neutron stars.
\end{abstract}

\maketitle

\section{Introduction}
\label{s:introduct}

The importance of Coulomb ionic mixtures cannot be understated in
many fields of physics and astrophysics. In astrophysics, dense
Coulomb plasmas are encountered in  neutron star crusts (e.g. Refs.\
\cite{Potekhin_etal97,CB03,CB04, CB10}), in white dwarfs (e.g.
Refs.\
\cite{IM85,Isern_etal91,BH01,DB02,Althaus_etal10,Garcia_etal10}), and in
giant planets (e.g. Ref.\ \cite{CSP06}). Similar properties possess
also dusty plasmas (e.g. Ref.\ \cite{Vaulina_etal10}) with numerous
applications in science and technology. The properties of dense
Coulomb plasmas are also important for inertial confinement fusion
(e.g. Ref.\ \cite{AS04}), antimatter (e.g. Ref.\
\cite{Andersen_etal10}), and ultra cold plasmas (e.g. Ref.\
\cite{Killian07}). Many applications of such plasmas involve
diffusion.

To describe ion diffusion in Coulomb plasmas one needs the
expressions for the diffusion currents and diffusion coefficients.
The first problem was addressed in our previous work \cite{BY13}.
The second problem is discussed here.

There is comprehensive astrophysical literature devoted to diffusion
of ions in dense stellar matter. The specific feature of this
diffusion is the long-ranged Coulomb interaction between ions. In
this respect the diffusion of ions has much in common with the
diffusion of particles interacting via a Yukawa potential with sufficiently large
screening length. The physics of diffusion has many aspects. One can study different types of diffusion coefficients. Most often considered are self-diffusion coefficients $D_\mathit{ii}$ and, somewhat less often,
but
more important, interdiffusion coefficients $D_\mathit{ij}$, which enter the expressions for the diffusion currents. Here, $i,j=1,2,\ldots$ enumerate ion species in a multicomponent plasma (MCP). In a one-component plasma (OCP) of ions there is only one self-diffusion coefficient $D_1$. Note that a self-diffusion coefficient $D_\mathit{ii}$ in MCP should not be confused with a self-diffusion coefficient $D_1$ in OCP. One can further consider weak or strong Coulomb coupling, classical or quantum ion motion, the presence of a magnetic field, degenerate or non-degenerate electrons, etc.  Diffusion is studied with different techniques such as the Chapman-Enskog approach, Green-Kubo relations, molecular dynamics (MD) simulations, and effective potential method, as well as other methods and their combinations. Some of these cases and methods are discussed below in more detail.

We mainly focus on the inter-diffusion of ions in binary ionic mixtures (BIMs) which form either Boltzmann gas or a strongly coupled Coulomb liquid. The ions are assumed to be fully ionized and the electrons strongly degenerate (although these restrictions are not very important). The diffusion in a gas is a classical issue, well studied and described in well-known monographs  \cite{CC52,Hirsh54}; the diffusion in liquid is less elaborated.  Our aim will be to provide a unified treatment of  the diffusion coefficients in
ion gas and liquid and to present the results in a form convenient for using in numerical simulations of ion diffusion and related phenomena. In a BIM, there is one independent interdiffusion coefficient $D_{12}=D_{21}$ and two self-diffusion coefficients $D_{11}$ and $D_{22}$.

Weak Coulomb coupling means that the ions constitute almost ideal
gas. They are moving more or less freely and diffuse due to
relatively weak Coulomb collisions with neighboring ions. The diffusion
coefficients in this limit are usually expressed through a Coulomb
logarithm $\Lambda$, which can be estimated as the logarithm of the large ratio of the maximum to minimum impact parameters of colliding ions. Calculations are done using the classical theory of
diffusion in rarefied gases (see Refs.\ \cite{CC52,Hirsh54}). In astrophysical literature, this theory is often called the Chapman-Salpeter theory (meaning the application of the general
theory to diffusion due to Coulomb interaction). Early astrophysical publications based on this theory are cited, for instance, in Ref.\ \cite{Paquette86}. One can further consider the classical and quantum limits in ion-ion scattering (note that the motion of ions is always classical at weak coupling, quantum effects can emerge only in scattering events). In the classical limit, the minimum impact parameter in the expression for $\Lambda$ is determined by the classical distance of the closest approach of colliding ions. In the quantum limit the minimum impact parameter is determined by the de Broglie wavelengths of ions. One can also consider the cases of non-degenerate and degenerate electrons. In the latter case the electrons produce much weaker screening of the Coulomb interaction (i.e., contribute much less to the maximum impact parameter) than in the former case. We will focus on the classical scattering limit in the presence of strongly degenerate electrons.

When the coupling becomes stronger, the ratio of the maximum to minimum impact parameters decreases reducing the Coulomb logarithm. At intermediate couplings the Coulomb logarithm becomes $\Lambda \sim 1$, and the diffusion coefficients $D \sim a^2 \omega_p$, where $a$ is a typical inter-ion distance and $\omega_p$ is the ion plasma frequency (see Sec.\ \ref{s:Diffusion coefficients}). Characteristic ion-ion collision frequencies become comparable to $\sim \omega_p$, and typical ion mean free paths are $\sim a$.

At strong coupling the ions are mostly confined (caged) in their local potential wells (within respective Wigner-Seitz cells) and constitute either a Coulomb liquid or Coulomb crystal. Thus, the ions mainly oscillate around (quasi-) equilibrium positions and diffuse through thermally activated jumps from one position to another (neighboring) one. The first experimental observations of the caging effect in relaxation of strongly coupled plasmas were made by \citet{Bannasch12}. Here one can distinguish the cases of classical (the temperature $T \gtrsim T_p$) and quantum ($T \lesssim T_p$) ion motion (where $T_p=\hbar \omega_p/k_B$ is the ion plasma temperature that is close to the Debye temperature, with $k_B$ being the Boltzmann constant). In the quantum case collective oscillations (plasmons) play an important role. As for electrons, one can study the cases of a rigid (incompressible) electron background or weakly polarizable background. The latter case
is similar to the case of
ions interacting via Yukawa potentials (with sufficiently large screening length). We consider the diffusion in Coulomb liquid neglecting quantum effects but taking into account both cases of rigid and slightly polarizable electron background. These cases give essentially the same results.

A semianalytic consideration of weak coupling was developed by
\citet{FM79in} who provided the expressions for $D_{\mathit{ij}}$
through a Coulomb logarithm and developed a computer code for
calculating $D_\mathit{ij}$. The authors considered the cases of
quantum and classical minimum impact parameters in the Coulomb
logarithm and introduced the resistance coefficients
$K_{\mathit{ij}}$ (that determine the ``friction forces'' inversely
proportional to $D_{\mathit{ij}}$). Their results were extended and
used by \citet{IM85} (in the case of weak coupling) who simulated
the evolution of $^{12}$C -- $^{16}$O white dwarfs.

\citet{Paquette86} calculated binary diffusion coefficients at weak
and moderate couplings using the Chapman-Enskog (Chapman-Spitzer)
formalism with a statically screened Coulomb potential. The authors
presented accurate analytic fits of collision integrals (tabulated
spline coefficients). Their results are applicable as long as
Coulomb coupling is not very strong. They discussed also earlier MD
calculations of the self-diffusion coefficient at strong coupling.

Pioneering MD calculations of the self-diffusion coefficient $D_1$
in OCP were performed
in 1975
by \citet{Hansen75}.
For the ion coupling
parameter $\Gamma>1$ (defined in Sec.\ \ref{s:Model}) they proposed
the fit
\begin{equation}
   D^*_1=D_1/(\omega_pa^2) \approx 2.95 \,\Gamma^{-4/3}.
\label{e:Hansen75}
\end{equation}
%

\citet{HJM85} carried out
MD calculations of $D_{12}$, $D_{11}$, and
$D_{22}$ in BIMs in the regime of intermediate and strong couplings.
They presented the approximate relation [their Eq.\ (23)] between the inter- and
self-diffusion coefficients,
\begin{equation}
 D_{12}\approx x_2 D_{11}+x_1 D_{22}.
\label{e:D12=D11+D22}
\end{equation}
They tabulated $D_{12}$, $D_{11}$, and $D_{22}$ for
some coupling strengths and relative fractions of ions ($x_1$ and
$x_2=1-x_1$) in the $^{1}$H -- $^{4}$He mixture.

\citet{BP87} performed MD and advanced kinetic theory calculations
of the interdiffusion coefficients in BIMs for strong and weak
couplings. The results were in good agreement with previous studies.
\citet{Robbins88} considered self-diffusion in OCP using MD of
Yukawa systems. \citet{RNZ95} performed MD calculations of BIMs for
wide ranges of $m_2/m_1$ and $Z_2/Z_1$ (ion mass and charge ratios)
at strong, moderate and weak
coupling in Coulomb plasmas and in Yukawa systems; they studied
self-diffusion and inter-diffusion, and emphasized close relation
between these systems and the systems of hard spheres.

\citet{OH00} did extensive MD calculations of the self-diffusion
coefficient in OCP Yukawa systems. They used the Green-Kubo
relation and the ordinary space diffusion formula to determine
$D_1$ (and the results converge). They tabulated the computed
values of $D_1^*=D_1/(\omega_pa^2)$ and approximated $D_1^*$ by the
expression
\begin{equation}
     D_1^*=\alpha (T^*-1)^\beta +\gamma,
\label{e:D1}
\end{equation}
where $T^*=T/T_m$, and $T_m$ is the melting temperature.
They presented the fit parameters $\alpha$,
$\beta$, and $\gamma$ as functions of the screening parameter in the
Yukawa potential and obtained good agreement with the results for
Coulomb systems in the cases of large screening lengths in the Yukawa potentials.

\citet{DM05} performed MD calculations of the self-diffusion
coefficient in OCP using a semiempirical potential and fitted
the results by Eq.\ (\ref{e:D1}) with $\gamma=0.028$,
$\alpha=0.00525$, and $\beta=1.154$. As the next step \citet{Daligault06}
analyzed liquid dynamics in a strongly coupled OCP and concluded
that although dynamical behavior of ions (with long-range Coulomb
interaction) at strong coupling changed from almost free particle
motion to the caging regime, the universal laws or ordinary liquids
with short-range interaction remained valid there.

\citet{Hughto10} performed MD calculations of the self-diffusion
coefficient of $^{22}$Ne in a mixture of many ion species at
strong coupling. They presented an original fit [their Eq.\ (8)] for
$D_{\mathit{ii}}/D_1$ (a combination of exponents and powers of
$\Gamma$).

In his next paper \citet{Daligault12} performed  MD simulations of self-diffusion in
OCP and BIMs at strong coupling and fitted the results by [his Eq.
(4)]
\begin{equation}
        D^*=\frac{D}{\omega_p a^2}= \frac{A}{\Gamma}\, \exp(-B \Gamma),
\label{e:caging+jumps}
\end{equation}
which would be appropriate to the regime of caging and thermally
activated jumps ($A$ and $B$ being some fit parameters). He
considers previous fits at strong coupling [like Eq.~\eqref{e:D1}]
as  less physical. In addition, he used standard Chapman-Spitzer
results at weak coupling and emphasized very good agreement of MD
and Chapman-Spitzer approaches at intermediate coupling. Later
\citet{Daligault12a} suggested similar ideas for Yukawa
OCP systems.

As the next step \citet{Khrapak13} considered the self-diffusion
coefficient in OCP. He used the standard Chapman-Spitzer theory at
weak coupling and results of MD calculations by different authors at
strong coupling. Based on those results he suggested a simple and
convenient analytic approximation, which reproduced the cases of weak
and strong couplings, by introducing a generalized Coulomb logarithm
$\Lambda_\mathrm{eff}$.

Finally, quite recently \citet{BD13} put forward the idea that the
cases of weak and strong coupling can be described within the same
formalism of the effective binary interaction potential and
traditional Chapman-Enskog theory (even at strong
Coulomb coupling!). They constructed some examples of the effective
potential inferred from radial distribution functions of ions $g(r)$;
these functions were computed via the
hyper-netted chain (HNC) approach. The effective potential allows
one not only to account for the screening effects (this can be done
by employing the screened Coulomb potential), but also take into account
even strong correlations between the ions. This method treats the screening and
correlation effects self-consistently; no ``external'' screening
lengths are involved. The authors compared the self-diffusion
coefficients in OCP calculated by different methods (their Fig.\ 2)
and emphasized the importance of expressing the diffusion
coefficients through generalized Coulomb logarithms. We will follow
this approach extending it to BIMs.

For the completeness of our consideration let us mention some others methods which have also been used to calculate diffusion coefficients in simulations of some phenomena in dense stars.

\citet{BH01} proposed to employ the self-diffusion coefficient $D_1$
to study $^{22}$Ne settling in white dwarfs (at strong coupling).
They tried two forms of $D_1$. First, they took $D_1$ using the
Stokes-Einstein relation for a particle of radius $a_p$ (taken to be
the radius of the ion sphere for $^{22}$Ne) moving in a fluid with
viscosity $\eta$ obtained from fits to the results of MD simulations. This
method is suitable for inter-diffusion of trace ions of one species
in BIMs. Second, the authors used $D_1$ obtained in Ref.\ \cite{Hansen75}.
They found that the values of $D_1$ estimated in these two ways
were close and led to the same results.

\citet{DB02} compared the same two different forms of self-diffusion
coefficients at strong coupling to study the $^{22}$Ne settling in
white dwarfs. In addition, they  took into account computational
uncertainties of $\eta$ and obtained that these uncertainties did not
affect noticeably $D_1$. They suggested using $D_1$ taken from Ref.\ \cite{Hansen75} in modeling diffusion processes.

\citet{P07} simulated sedimentation and x-ray bursts in neutron
stars. In their Appendix they described the
resistance coefficients and associated diffusion coefficients. They
proposed piece-like interpolation of weak coupling and strong coupling
cases. They  considered weak coupling following \citet{FM79} and
strong coupling following Ref.\ \cite{Hansen75}.

Although we do not study diffusion in Coulomb crystals let us
mention that the problem was investigated by \citet{Hughto11} using
MD with the natural result that this diffusion is strongly
suppressed in comparison with that in Coulomb liquid.

It is also worth to mention some papers devoted to diffusion in
magnetized Coulomb plasmas. For instance, \citet{Bernu81} calculated
the self-diffusion coefficient in OCP with a constant uniform magnetic
field $\bm{B}$. Much later \citet{RJW03} repeated MD calculations of
self-diffusion in OCP in a magnetic field. They obtained two
self-diffusion coefficients, $D_\parallel$ and $D_\perp$, along and
across $\bm{B}$. Both coefficients decrease with increasing $B$,
and $D_\perp < D_\parallel$.

\section{HNC calculation of effective potential}
\label{s:Model}

Consider a classical (quantum effects neglected) non-magnetized
binary ionic mixture (BIM), which consists of two ion species and
neutralizing rigid electron background \cite{HTV77}. An assumption
of the rigid electron background allows us to factorize out the
electrons while calculating inter-ionic diffusion \cite{BY13}. Let
$n_j$, $A_j$, and $Z_j$ be, respectively, the number density, mass, and charge numbers of
ion species $j=1$ and 2. For certainty, we set
$Z_1<Z_2$. Let $n=n_1+n_2$ denote the overall ion number density and
$x_j={n_j}/{n}$ the fractional number of ions $j$
(with $x_1+x_2=1$). Then we can define the mean value $\overline{f}$
of any quantity $f_j$ in a BIM as $\overline{f}=x_1 f_1 + x_2 f_2$. In
the following (unless the contrary is indicated)
lengths are measured in the units of the ion-sphere radius,
\begin{equation}
   a = \left( \frac{3}{4 \pi n} \right)^{\frac{1}{3}},
\label{e:a}
\end{equation}
and all potentials in units of $\slfrac{k_B T}{e}$
($e$ being the elementary charge).

A state of the BIM is defined by ion charge and mass numbers and by
two dimensionless parameters, the  fractional number $x\equiv x_1$
of ions 1, and the Coulomb coupling parameter $\Gamma_0$ (see Refs.\
\cite{HTV77,HJM85}),
\begin{equation}
   \Gamma_0 = \frac{e^2}{a k_B T}.
\label{e:Gamma}
\end{equation}
We can also introduce the Coulomb coupling parameter for each ion species
(see, e.g., Ref.\ \cite{HPY07}),
\begin{equation}
   \Gamma_j = \frac{Z_j^2 e^2}{a_j k_B T} = \frac{Z_j^{\frac{5}{3}} e^2}{a_e k_B T},
\label{e:Gammaj}
\end{equation}
where $a_e = (\slfrac{3}{4\pi n_e})^{\slfrac{1}{3}}$ is the
electron-sphere radius, $a_j = a_e Z_j^{\slfrac{1}{3}}$ is the ion
sphere radius of species $j$, and $n_e = Z_1 n_1 + Z_2 n_2 =
\overline{Z}n$ is the electron number density.

Furthermore, it is convenient to introduce the mean ion coupling parameter
$\overline{\Gamma}=x_1 \Gamma_1 + x_2 \Gamma_2$, which can be expressed as
\begin{equation}
   \overline{\Gamma} = \Gamma_0 \overline{Z^{\frac{5}{3}}}\overline{Z}^{\frac{1}{3}},
\label{e:GammaAver}
\end{equation}
and which reduces to $\overline{\Gamma}=\Gamma_0 Z^2$ in the case of OCP.

Let $g_\mathit{ij}(r)$, $h_\mathit{ij}(r)$, and $c_\mathit{ij}(r)$
($i,j=1,2$) be the radial distribution functions (RDFs), the total and direct
correlation functions, respectively (as detailed, e.g., in Ref.\
\cite{Croxton74}). All these functions are symmetric [i.e.
$g_\mathit{ij}(r)=g_\mathit{ji}(r)$], and
$h_\mathit{ij}(r)=g_\mathit{ij}(r)-1$. The effective potential
$\Phi(r)$ in OCP is introduced by the relation  $g(r)=\exp{[-\Phi(r)]}$
\cite{BD13,Croxton74}. The extension of this relation to the BIM
case is straightforward,
\begin{equation}
   g_\mathit{ij}(r)=\exp[-\Phi_\mathit{ij}(r)].
\label{e:RDF-phi-TCP}
\end{equation}
One primarily needs $\Phi_{12}(r)$ for calculating the
interdiffusion coefficient.

Generally, all these functions
cannot be calculated analytically. We calculate them by the HNC
method, which is known to be sufficiently accurate (as detailed in Sec.\ \ref{s:Discussion}) and relatively simple (e.g., Refs.\ \cite{HTV77,SPS73,Ng74}).
Let us outline this method to simplify the reading of this paper.
It consists in solving  together the equations of two types,
the Ornstein-Zernike equations relating direct and total correlation
functions and the HNC closure relations. Since the equations are
used in Fourier space, we define the dimensionless Fourier transform
as
\begin{equation}
   \hat{f}(k) = \frac{4 \pi}{k} \int_0^{+ \infty}\! f(r)r \sin{(kr)} \, dr
\label{e:Fourier}
\end{equation}
(wave number $k$ is measured in units of $\slfrac{1}{a}$), and its
inverse as
\begin{equation}
   f(r) = \frac{1}{2 \pi^2 r} \int_0^{+ \infty}\! \hat{f}(k)k \sin{(kr)} \, dk.
\label{e:InverseFourier}
\end{equation}
Then the Ornstein-Zernike relations are readily written as
\cite{HTV77},
\begin{equation}
   \hat{h}_\mathit{ij}(k) = \hat{c}_\mathit{ij}(k) + \frac{3}{4 \pi} \sum_{q=1}^{2} x_q \hat{h}_\mathit{iq}(k)
   \hat{c}_\mathit{qj}(k),
\label{e:OZ-base}
\end{equation}
and the HNC closure is
\begin{equation}
   g_\mathit{ij}(r) =  h_\mathit{ij}(r) + 1 = \exp[h_\mathit{ij}(r) - c_\mathit{ij}(r) -
   \phi_\mathit{ij}(r)],
\label{e:HNC-closure-base}
\end{equation}
$ \phi_\mathit{ij}(r)$ being the bare Coulomb interaction,
\begin{equation}
   \phi_\mathit{ij}(r) = \frac{Z_i Z_j \Gamma_0}{r}.
\label{e:bare-Coulomb}
\end{equation}

Equations \eqref{e:OZ-base} and \eqref{e:HNC-closure-base} form a closed
set of six equations for $h_\mathit{ij}$ and $c_\mathit{ij}$, but
they cannot be solved directly due to the long-range nature of the Coulomb
potential. For OCP this problem was circumvented by Springer
et al.\ \cite{SPS73} and Ng \cite{Ng74} by introducing
short-ranged potentials and correlation functions. A similar method
was used by Hansen et al.\ \cite{HTV77} for BIMs. Let us
outline this method here for the sake of
completeness.

In our case the total correlation functions $h_\mathit{ij}(r)$ are
short-ranged and the direct correlation functions have the
asymptotes \cite{HTV77,SPS73,Ng74}
\begin{equation}
   \lim_{r \to \infty} c_\mathit{ij}(r)  =  - \phi_\mathit{ij}(r).
\label{e:c-asympt}
\end{equation}
Let us introduce a quantity
\begin{equation}
   \gamma_{\mathit{ij}}(r) = h_{\mathit{ij}}(r) - c_{\mathit{ij}}(r)
\label{e:gamma-def}
\end{equation}
which has the asymptotic property
\begin{equation}
   \lim_{r \to \infty} \gamma_{\mathit{ij}}(r)  = \phi_{\mathit{ij}}(r).
\label{e:gamma-asympt}
\end{equation}
Then we define the short-range $(s)$ correlation functions and
potentials,
\begin{align}
   \gamma^{(s)}_{\mathit{ij}}(r)  & =   \gamma_{\mathit{ij}}(r) - \phi^{(l)}_{\mathit{ij}}(r),
\label{e:gamma-s} \\
   c^{(s)}_{\mathit{ij}}(r)  & =   c_{ij}(r) + \phi^{(l)}_{\mathit{ij}}(r),
\label{e:c-s} \\
   \phi^{(s)}_{\mathit{ij}}(r) & = \phi_{\mathit{ij}}(r) - \phi^{(l)}_{\mathit{ij}}(r).
\label{e:phi-s}
\end{align}
The long-range $(l)$ functions $\phi^{(l)}_{\mathit{ij}}(r)$ have to satisfy
two conditions, (1) possess the same asymptotes as $\phi_{\mathit{ij}}(r)$ at
$r \to \infty$ and (2) be regular at $r=0$. Otherwise, they are
arbitrary. Following Ng \cite{Ng74}, we choose
\begin{equation}
   \phi^{(l)}_\mathit{ij}(r) = \frac{Z_i Z_j \Gamma_0}{r}\, \mathrm{erf}(\alpha
   r),
\label{e:phi-L}
\end{equation}
with $\alpha=1.1$; its Fourier transform in Eq.\ \eqref{e:OZ-base}
is
\begin{equation}
   \hat{\phi}^{(l)}_\mathit{ij}(k) = \frac{4 \pi Z_i Z_j \Gamma_0}{k^2}\, \exp \left(-\frac{k^2}{4
   \alpha^2}\right).
\label{e:Fphi-L}
\end{equation}

Now we rewrite Eqs.\ \eqref{e:OZ-base} and
\eqref{e:HNC-closure-base} in terms of short-ranged correlation
functions and potentials,
\begin{align}
&  \hat{\gamma}^{(s)}_\mathit{ij}(k) +
\hat{\phi}^{(l)}_\mathit{ij}(k) = \frac{3}{4 \pi} \sum_{q=1}^{2}
x_q\, \left[\hat{\gamma}^{(s)}_\mathit{iq}(k) +
\hat{c}^{(s)}_\mathit{iq}(k)\right]
\nonumber \\
&\times \left[\hat{c}^{(s)}_\mathit{qj}(k) -
\hat{\phi}^{(l)}_\mathit{qj}(k) \right] ,
\label{e:OZ-final} \\
&   g_\mathit{ij}(r)  = \exp [\gamma^{(s)}_\mathit{ij}(r) -
\phi^{(s)}_\mathit{ij}(r)],
\label{e:HNC-final} \\
&   c^{(s)}_\mathit{ij}(r)  = g_\mathit{ij}(r) -
\gamma^{(s)}_\mathit{ij}(r) - 1. \label{e:Cs-final}
\end{align}
%
%
This system can be solved iteratively starting with a guess for
$c^{(s)}_{\mathit{ij}}(r)$. Before that the functions
$\hat{\gamma}^{(s)}_{\mathit{ij}}(k)$ should be explicitly expressed from
Eqs.\ \eqref{e:OZ-final}. As Eqs.\ \eqref{e:OZ-final} are linear
with respect to $\hat{\gamma}^{(s)}_{\mathit{ij}}(k)$, they can be solved
analytically once and for all. We will not write here the resulting
formulas because they are inconveniently large and their derivation
is obvious. Points $k = 0$ and $r = 0$ require special consideration
because these values cannot be substituted in Eqs.\ \eqref
{e:OZ-final} and \eqref{e:HNC-final} due to singularities in
$\hat{\phi}^{(l)}_\mathit{ij}(k)$ and $\phi^{(s)}_\mathit{ij}(r)$,
respectively. The problem is dealt with as following. First, the
values of $\hat{\gamma}^{(s)}_\mathit{ij}(0)$ and $g_\mathit{ij}(0)$
are calculated separately,
\begin{align}
   &g_\mathit{ij}(0) = 0, \,
   \hat{\gamma}^{(s)}_{11}(0) = -\frac{4 \pi}{3 x_1} - \hat{c}^{(s)}_{11}(0), \nonumber \\
  &\hat{\gamma}^{(s)}_{12}(0) = - \hat{c}^{(s)}_{12}(0), \quad
   \hat{\gamma}^{(s)}_{22}(0) = -\frac{4 \pi}{3 x_2} - \hat{c}^{(s)}_{22}(0)
\label{e:zero-val}
\end{align}
[$\hat{\gamma}^{(s)}_{\mathit{ij}}(0)$ being a $k \to 0$ limit of the
solutions of Eqs.\ \eqref{e:OZ-final}]. Second, Fourier and inverse
Fourier transforms are rewritten to handle
\begin{align}
   &\hat{f}(0) = 4 \pi \int_0^{+ \infty}\! f(r)r^2 \, dr, \nonumber \\
   &f(0) = \frac{1}{2 \pi^2} \int_0^{+ \infty}\! \hat{f}(k)k^2 \, dk.
\label{e:Fourier-zero}
\end{align}

Numerical calculations were performed on a mesh of
$N_{p}=2049$ points running from 0 to $r_{\mathrm{max}}$
and from 0 to $k_{\mathrm{max}}$; $r_{\mathrm{max}}$ was taken to be
80 (other values for $N_{p}$ and $r_{\mathrm{max}}$ were
also taken to check the stability of numerical procedures);
$k_{\mathrm{max}}$ was computed from the standard relation
\begin{align}
   &\Delta r = \frac{r_{\mathrm{max}}}{N_{p} - 1}, \,
   \Delta k = \frac{\pi}{\left( N_{p}-1 \right) \Delta r}, \nonumber \\
   &k_{\mathrm{max}} = \left(N_{p}-1 \right) \Delta k.
\label{e:Mesh}
\end{align}
Fourier integrals were discretized on a mesh using Simpson's rule
(intermediate points were calculated via cubic spline interpolation)
and processed by means of appropriate fast Fourier transform. A convergence criterion for iterative process was taken to be
\begin{equation}
   \sqrt{\int_0^{r_{\mathrm{max}}} \! \left( g_{22}^{(q)}(r) - g_{22}^{(q-1)}(r) \right)^2 \, dr} < 10^{-7},
\label{e:Convergence}
\end{equation}
because $g_{22}$ converges slower than  $g_{12}$ or  $g_{11}$ (here $q$ is the iteration number).

After the computations have been completed, the accuracy of our results
has been checked by comparing the excess (Coulomb) potential energy with
the results of Ref.\ \cite{HTV77}. The agreement has been found to be quite satisfactory (energies have been reproduced up to five to six significant digits).

The examples of HNC results for a mixture of $^{1}$H and $^{12}$C
($x_1=0.3$, $x_2=0.7$) are presented in Figs.\ 1 and 2.
\begin{figure}[t]
\centering
\includegraphics[viewport = 0 0 229 203,width=0.55\textwidth]{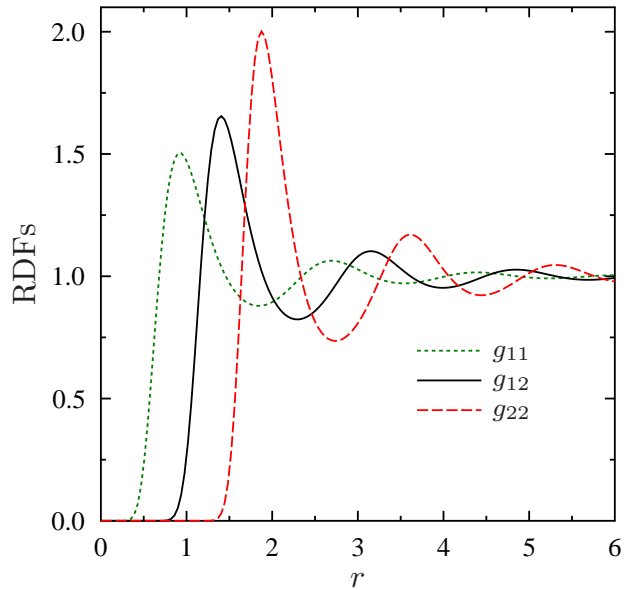}
\caption{(Color online) Radial distribution functions for a mixture
composed of 30\% $^{1}$H and 70\% $^{12}$C (by numbers), with
$\Gamma_0 = 5\,\,(\overline{\Gamma}\approx 117)$.}
\label{fig:rdf}
\end{figure}
\begin{figure}[t]
\centering
\includegraphics[viewport = 0 0 229 203,width=0.55\textwidth]{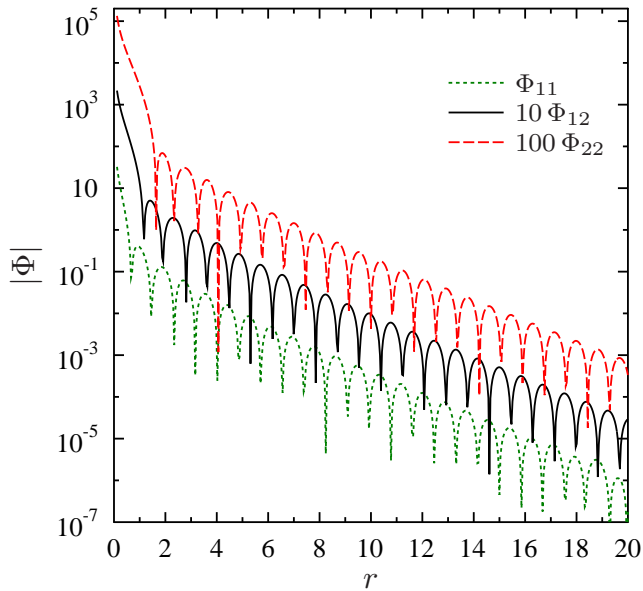}
\caption{(Color online) Absolute values of effective potentials for
the same mixture of $^{1}$H and $^{12}$C as in Fig.\ \ref{fig:rdf}.
For a better visualization, $\Phi_{12}$ is multiplied by 10 and
$\Phi_{22}$ by 100.} \label{fig:phi}
\end{figure}
%


\section{Diffusion coefficients}
\label{s:Diffusion coefficients}

\renewcommand{\arraystretch}{1.45}
\setlength{\tabcolsep}{4pt}
\begin{table*}[tb]
\centering \caption{Fit parameters in Eq. \eqref{e:Lambda-fit} as
well as rms and maximum fit errors. The last column contains values of $x_1$ and $\Gamma_0$ at which maximum fit error is achieved}
    \begin{tabular}{l c c c c >{\centering\arraybackslash}p{1.2cm} c c l}
    \toprule
    \hline
    \multicolumn{1}{ c }{Mixture} &  $p_1$  &  $p_2$  &  $p_3$  &  $p_4$  &  $p_5$  &  $\delta_\mathrm{rms}$, \% & $\delta_\mathrm{max}$, \%  &  \multicolumn{1}{ c }{$(x_1, \Gamma_0)_{\mathrm{max}}$}\\ \hline
    \midrule
    $^{1}$H -- $^{4}$He & $7.43\!\times\! 10^{-2}$ & $-1.13\!\times\! 10^{-2}$ & $1.72\!\times\! 10^{-1}$ & $8.57\!\times\! 10^{-2}$ & $1.45$ & $3.1$ & $10$ & $(0.7,0.4)$\\
    $^{1}$H -- $^{12}$C & $3.80\!\times\! 10^{-2}$ & $6.57\!\times\! 10^{-3}$ & $2.52\!\times\! 10^{-2}$ & $1.39\!\times\! 10^{-1}$ & $1.34$ & $5.6$ & $18$ & $(0.99,0.729)$\\
    $^{4}$He -- $^{12}$C & $7.01\!\times\! 10^{-3}$ & $9.08\!\times\! 10^{-4}$ & $1.09\!\times\! 10^{-2}$ & $1.17\!\times\! 10^{-1}$ & $1.41$ & $4.0$ & $13$ & $(0.9,5.785)$\\
    $^{12}$C -- $^{16}$O & $9.95\!\times\! 10^{-5}$ & $-6.35\!\times\! 10^{-6}$ & $1.61\!\times\! 10^{-3}$ & $3.96\!\times\! 10^{-2}$ & $1.48$ & $2.6$ & $10$ & $(0.9,0.015)$\\
    $^{16}$O -- $^{79}$Se & $7.22\!\times\! 10^{-5}$ & $5.00\!\times\! 10^{-5}$ & $1.14\!\times\! 10^{-4}$ & $1.33\!\times\! 10^{-1}$ & $1.38$ & $4.1$ & $16$ & $(0.9,0.187)$\\
    \hline
    \end{tabular}
\label{Tab:FitParam}
\end{table*}
\setlength{\tabcolsep}{6pt}
\renewcommand{\arraystretch}{1.0}

The standard Chapman-Enskog procedure gives the following leading
order approximation to the interdiffusion coefficient in a binary
mixture \cite{CC52,Hirsh54} (here in ordinary CGS units):
\begin{equation}
   D_{12} = \frac{3}{16} \frac{k_B T}{\mu n}
   \frac{1}{\widetilde{\Omega}_{12}^{(1,1)}},
\label{e:D-general-def}
\end{equation}
where $\mu = \slfrac{m_1 m_2}{(m_1+m_2)}$ is a reduced mass of
colliding ions, and $\widetilde{\Omega}$ is a collisional integral
defined below. The second order approximation to $D_{12}$ will be
outlined in the next section.

Let us introduce ``hydrodynamic'' plasma frequency for a mixture
(e.g., Ref.\ \cite{HJM85}):
\begin{equation}
   \omega_{p} = \sqrt{\frac{4 \pi n \overline{Z}^2
   e^2}{\overline{A}m_0}},
\label{e:wp}
\end{equation}
$m_0$ being the atomic mass unit. Let us express the interdiffusion
coefficient in units of $\omega_{p} a^2$ through a dimensionless
collisional integral,
\begin{equation}
   D_{12}^{*} = \frac{D_{12}}{\omega_{p} a^2} =
   \frac{\pi^{\frac{3}{2}}}{2 \sqrt{6}} \frac{1}{\sqrt{\Gamma_0}}
   \sqrt{\frac{\overline{A}\left(A_1+A_2 \right)}{\overline{Z}^2 A_1 A_2}}
   \frac{1}{\Omega_{12}^{(1,1)}}.
\label{e:D-dim-def}
\end{equation}
Dimensionless collisional integrals are defined as (see, e.g.,
Refs.\ \cite{Paquette86} and  \cite{LL_Mechanics76})
\begin{align}
&   \Omega_{12}^{(\xi,\zeta)} = \int_0^{\infty}\!
\exp(-y^2)\,y^{2\zeta+3}Q_{12}^{(\xi)}(y) \, dy,
\label{e:Omega-dim-def}\\
&  Q_{12}^{(\xi)}(u) = 2 \pi \int_0^{\infty} \! \left[1 -
\cos^\xi{\left(\chi_{12}(b,u)\right)} \right]b\, db,
\label{e:Q-dim-def}\\
&  \chi_{12}(b,u) = \left| \pi- 2b
\int_{r_{12}^{\mathrm{min}}}^{\infty} \! \frac{dr}{r^2
\sqrt{1-\frac{b^2}{r^2}-\frac{\phi_{12}}{u^2}}} \right|,
\label{e:chi-dim-def}
\end{align}
where $\chi_{12}$ is the classical scattering angle, $b$ is the
impact parameter, $\phi_{12}$ the interaction potential between
particles $1$ and $2$, $u$ is the dimensionless relative velocity
(at infinity; in units of $\sqrt{\slfrac{2 k_B T}{\mu}}$),
$r_{12}^{\mathrm{min}}$ is the distance of the closest approach
[i.e. maximum root of the denominator in the integrand
\eqref{e:chi-dim-def}].

For a weakly coupled (WC) BIM ($\overline{\Gamma} \ll 1$), the
diffusion coefficient \eqref{e:D-dim-def} can be calculated
analytically (e.g., Refs.\ \cite{CC52,Hirsh54}):
\begin{equation}
   D_{12}^{*\mathrm{(WC)}} =  \sqrt{\frac{\pi}{6}}\,
   \frac{1}{\Gamma_0^{\frac{5}{2}}}
   \sqrt{\frac{\overline{A}\left(A_1+A_2 \right)}{\overline{Z}^2 A_1
   A_2}}\,
   \frac{1}{Z_1^2 Z_2^2 \Lambda^{\mathrm{(WC)}}},
\label{e:D-WC-def}
\end{equation}
where $\Lambda^{\mathrm{(WC)}}$ is a ``classical'' Coulomb logarithm
for a weakly coupled plasma,
\begin{equation}
   \Lambda^{\mathrm{(WC)}} = \ln{\left( \frac{1}{\sqrt{3}\Gamma_0^{\frac{3}{2}}Z_1 Z_2 \sqrt{\overline{Z^2}}} \right)}.
\label{e:Lambda-WC-def}
\end{equation}
%

\renewcommand{\arraystretch}{1.45}
\setlength{\tabcolsep}{4pt}
\begin{table*}[tb]
\centering \caption{$\Gamma_0$ mesh points used for computing and
fitting $\Lambda_\mathrm{eff}$. For each BIM the points have been
distributed within three ranges I, II and III; $\Delta^{+}$ and
$\Delta^\times$ determine the distances between neighboring points
as described in the text. Lower bounds of each range are exact,
upper bounds are rounded up}
    \begin{tabular}{l l l l}
    \toprule
    \hline
    \multicolumn{1}{c}{Mixture}  &  $\Gamma_0$ range I   &  $\Gamma_0$ range II  &  $\Gamma_0$ range III  \\ \hline
    \midrule
    $^{1}$H -- $^{4}$He    & $[10^{-4},0.05],\Delta^{+}\! =\! 0.002$ & $[0.4,1.6],\Delta^{\times}\! =\!1.25$ & $[1.7,52],\Delta^{\times}\! =\! 1.3$\\
    $^{1}$H -- $^{12}$C    & $[10^{-4},0.01],\Delta^{+}\! =\!0.001$ & $[0.15,0.4],\Delta^{\times}\! =\!1.2$ & $[0.4,6],\Delta^{\times}\! =\! 1.35$\\
    $^{4}$He -- $^{12}$C   & $[10^{-4},0.005],\Delta^{+}\! =\!3.5\!\times\!10^{-4}$ & $[0.06,0.2],\Delta^{\times}\! =\!1.25$ & $[0.2,5.8],\Delta^{\times}\! =\! 1.4$\\
    $^{12}$C -- $^{16}$O   & $[10^{-4},0.003],\Delta^{+}\! =\!10^{-4}$ & $[0.015,0.05],\Delta^{\times}\! =\!1.35$ & $[0.055,3.2],\Delta^{\times}\! =\! 1.4$\\
    $^{16}$O -- $^{79}$Se  & $[10^{-5},2.5\!\times\!10^{-4}],\Delta^{+}\! =\!10^{-5}$ & $[0.003,0.01],\Delta^{\times}\! =\!1.22$ & $[0.01,0.2],\Delta^{\times}\! =\! 1.34$\\
    \hline
    \end{tabular}
\label{Tab:Gamma0Val}
\end{table*}
\setlength{\tabcolsep}{6pt}
\renewcommand{\arraystretch}{1.0}

Now the algorithm for computing $D_{12}^*$ at arbitrary coupling is
straightforward. First, we calculate RDFs using HNC method described
in Sec.\ \ref{s:Model}. Second, we find effective potential
$\Phi_{12}$ from Eq.\ \eqref{e:RDF-phi-TCP} and substitute it
instead of $\phi_{12}$ in the integral \eqref{e:chi-dim-def}. Then
we calculate $D_{12}^*$ from Eqs.\ \eqref{e:D-dim-def},
\eqref{e:Omega-dim-def}, and \eqref{e:Q-dim-def}.

\begin{figure}[t]
\centering
\includegraphics[viewport = 0 0 229 203,width=0.55\textwidth]{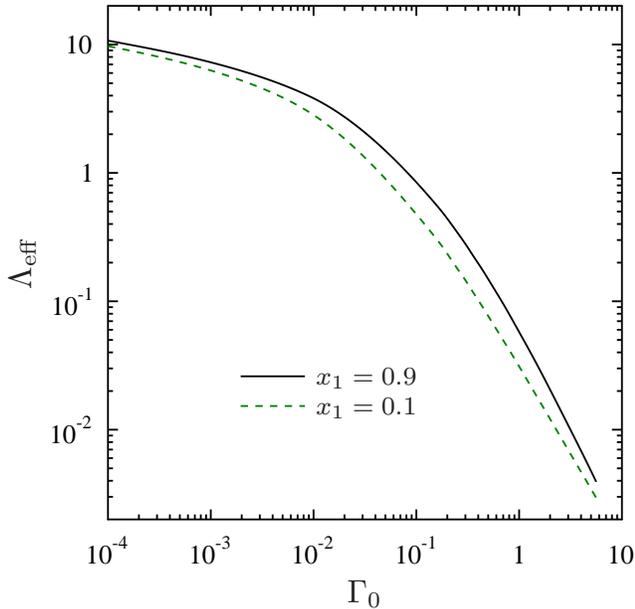}
\caption{(Color online) Effective Coulomb logarithm $\Lambda_\mathrm{eff}$ calculated from Eq.\ \eqref{e:Lambda-eff-def} for the $^{1}$H -- $^{12}$C mixture. Regions of weak and strong couplings can be clearly seen as well as a transition region between them.}
\label{fig:Lambda}
\end{figure}

We have performed such calculations of the interdiffusion
coefficients for $^{1}$H -- $^{4}$He, $^{1}$H -- $^{12}$C, $^{4}$He
-- $^{12}$C, $^{12}$C -- $^{16}$O, and $^{16}$O -- $^{79}$Se mixtures
for a variety of values of $\Gamma_0$ and $x_1$. We could have easily
considered other BIMs if necessary. The easiest way to present these
data is to fit the effective Coulomb logarithm by an analytic
expression. We have calculated $D_{12}^{*}$ and then
$\Lambda_{\mathrm{eff}}$ using the expression:
\begin{equation}
   \Lambda_{\mathrm{eff}} = \sqrt{\frac{\pi}{6}} \, \frac{1}{D_{12}^*
   Z_1^2 Z_2^2 \Gamma_0^{\frac{5}{2}}} \, \sqrt{\frac{\overline{A}\left(A_1+A_2 \right)}{\overline{Z}^2 A_1 A_2}}.
\label{e:Lambda-eff-def}
\end{equation}
Thus, $\Lambda_{\mathrm{eff}}$ coincides with
$\Lambda^{\mathrm{(WC)}}$, Eq.\ \eqref{e:D-WC-def}, in the weak
coupling limit. The examples of $\Lambda_{\mathrm{eff}}$ for $^{1}$H
-- $^{12}$C mixture are presented in Fig.\ \ref{fig:Lambda}.

Fitting $\Lambda_{\mathrm{eff}}$ instead of $D_{12}^{*}$ is more
convenient because $\Lambda_{\mathrm{eff}}$ is expected to be
relatively weakly dependent on plasma parameters (particularly on relative number density $x_1$). We propose the fit
\begin{equation}
   \Lambda_{\mathrm{eff}}\left( \Gamma_0, x_1 \right) = \ln{\left(1+\frac{p_1 x_1^2+p_2 x_2^2+p_3}
   {\Gamma_0^{p_4 x_1+p_5}} \right)},
\label{e:Lambda-fit}
\end{equation}
which contains five parameters $p_1,\ldots p_5$.
These parameters are presented in Table \ref{Tab:FitParam} along
with the root mean square (rms) relative deviation,
$\delta_\mathrm{rms}$, and the maximum relative fit errors,
$\delta_\mathrm{max}$. The $x_1$ mesh points have been taken as $x_1
= 0.01,0.1,0.2,0.3,\dots,0.9,0.99$. The $\Gamma_0$ mesh points have
been selected differently for each BIM (Table \ref{Tab:Gamma0Val}).
For each BIM, the mesh points have been distributed over three
ranges of $\Gamma_0$ labeled as I, II, and III in Table
\ref{Tab:Gamma0Val}. These ranges refer to weak, intermediate, and
strong Coulomb pairing, respectively (note that the actual strength
of Coulomb coupling is determined by $\overline{\Gamma}$, not by
$\Gamma_0$). In range I the points have been taken equidistant (any
next point being larger than the previous one by $\Delta^+$), whereas
in ranges II and III logarithmically equidistant (any next point was
$\Delta^\times$ times higher than the previous one).

\section{Discussion}
\label{s:Discussion}

\renewcommand{\arraystretch}{1.45}
\setlength{\tabcolsep}{4pt}
\begin{table*}[tb]
\caption{Comparison of $D^*_{12}$ calculated here with MD data and verification of Eq.\
\eqref{e:D12=D1+D2} for $^{1}$H--$^{4}$He and $^{1}$H--$^{12}$C mixtures.
The values of $D_{12}^{*\mathrm{MD}}$ are
taken from MD simulations of \citet{HJM85}. The values $D_{12}^{*\mathrm{S}}$ are
obtained from Eq.\ \eqref{e:D12=D1+D2} with the self-diffusion
coefficients calculated by the
effective potential method.}
    \begin{tabular}{l l l l l | l l l l}
    \toprule
    \hline
    \multicolumn{5}{c|}{$^{1}$H -- $^{4}$He}  & \multicolumn{4}{c}{$^{1}$H -- $^{12}$C}\\ \hline
    $x_1$  &  $\Gamma_0$  &  $D_{12}^*$  &  $D_{12}^{*\mathrm{S}}$  &  $D_{12}^{*\mathrm{MD}}$  & $x_1$  &  $\Gamma_0$  &  $D_{12}^*$  &  $D_{12}^{*\mathrm{S}}$ \\ \hline
    \midrule
    0.5    &  0.397    &  4.20      &   3.73     &  3.00     &  0.2  &  5.75  &  0.0572  &  0.0322 \\
    0.5    &  3.992    &  0.268    &  0.230    &  0.142   &  0.5  &  5.75  &  0.0635  &  0.0354 \\
    0.5    &  39.738  &  0.0290  &  0.0242  &  0.0109 &  0.8  &  5.75  &  0.0688  &  0.0445 \\
    0.75  &  40.831  &  0.0279  &  0.0235  &  0.0122 \\
    0.25  &  40.610  &  0.0277  &  0.0237  &  0.0076 \\
    \hline
    \end{tabular}
\label{Tab:Compar}
\end{table*}
\setlength{\tabcolsep}{6pt}
\renewcommand{\arraystretch}{1.0}

Before discussing the results let us make a few remarks.

(1) There is no strict proof for the existence of an effective pair
interaction potential which would {\em{entirely}} incorporate all
many-body effects (correlations) between particles in a strongly
coupled Coulomb plasma. Moreover, it seems highly unlikely that such a
potential could exist in principle. Nevertheless, the effective
potential method seems to be a promising tool for obtaining
reasonably accurate solutions of some problems
of strongly coupled dense plasmas (see the original
work by Baalrud and Daligault \cite{BD13}).

(2) We use a standard HNC procedure to calculate RDFs. Although some
improved HNC techniques have been developed (e.g. Ref.\
\cite{II83}), we consider the accuracy of the standard HNC method
sufficient for our purpose. As seen from Fig.\ 2 of Ref.\
\cite{BD13}, even using the ``exact'' RDFs computed via MD simulations
makes almost negligible changes to the resulting Chapman-Enskog
diffusion coefficient compared to using RDFs obtained via
standard HNC method.

(3) Second order Chapman-Enskog approximation to the interdiffusion
coefficient in a BIM can be written as  \cite{CC52, Hirsh54}
\begin{equation}
   \left[ D_{12}^{*} \right]_2 = \frac{D_{12}^{*}}{1-\Delta},
\label{e:SecondApprox}
\end{equation}
where
\begin{align}
 & \Delta = 5 (\mathrm{C}-1)^2 \frac{\mathrm{P_1} \frac{x_1}{x_2} + \mathrm{P_2}
 \frac{x_2}{x_1}+\mathrm{P_{12}}}{\mathrm{Q_1} \frac{x_1}{x_2}+\mathrm{Q_2}
 \frac{x_2}{x_1}+\mathrm{Q_{12}}},
\label{e:Delta}\\
 & \mathrm{P_1} = \left( \frac{A_1}{A_1+A_2} \right)^3 \mathrm{E_1},  \quad
\mathrm{P_2} = \left( \frac{A_2}{A_1+A_2} \right)^3 \mathrm{E_2},
\label{e:Delta-P}\\
 &\mathrm{P_{12}} =
 \frac{3(A_1-A_2)^2+4A_1A_2\mathrm{A}}{(A_1+A_2)^2},
\label{e:Delta-P12}\\
 &\mathrm{Q_1} = A_1 \mathrm{E_1} \frac{6 A_2^2+5A_1^2-4A_1^2\mathrm{B}+8A_1A_2\mathrm{A}}{(A_1+A_2)^3},
\label{e:Delta-Q}\\
 &\mathrm{Q_{12}} = \frac{3(A_1-A_2)^2(5-4\mathrm{B})+4A_1A_2\mathrm{A}(11-4\mathrm{B})}{(A_1+A_2)^2} \nonumber \\
&+\frac{2\mathrm{E_1}\mathrm{E_2}A_1A_2}{(A_1+A_2)^2},
\label{e:Delta-Q12}
\end{align}
\begin{align}
 &\mathrm{A} = \frac{\Omega_{12}^{(2,2)}}{5\Omega_{12}^{(1,1)}}, \
\mathrm{B} = \frac{5\Omega_{12}^{(1,2)}-\Omega_{12}^{(1,3)}}{5\Omega_{12}^{(1,1)}}, \
\mathrm{C} = \frac{2\Omega_{12}^{(1,2)}}{5\Omega_{12}^{(1,1)}},
\label{e:Delta-A-B-C}\\
 &\mathrm{E}_j = \frac{\Omega_{\mathit{jj}}^{(2,2)}}{5\Omega_{12}^{(1,1)}}
 \frac{(A_1+A_2)^2}{A_1A_2} \sqrt{\frac{2 A_1 A_2}{A_j (A_1+A_2)}},\
 j=1,2;
\label{e:Delta-E}
\end{align}
$\mathrm{Q_2}$ is obtained from $\mathrm{Q_1}$ by interchanging
indices 1~and 2. Integrals $\Omega_{\mathit{jj}}^{(2,2)}$ are
defined in exactly the same way as $\Omega_{12}^{(\xi,\zeta)}$ but
with $\Phi_{\mathit{jj}}$ instead of $\Phi_{12}$. We have performed
calculations of the second-order corrections and found that they
do not exceed 5\% for the $^1$H -- $^{12}$C mixture. For mixtures
of more similar ions these corrections are even smaller.
Consequently, we have neglected them as the accuracy of the results
is limited by the fit errors and by the effective potential method
itself.

\begin{figure}[t]
\centering
\includegraphics[viewport = 0 0 229 203,width=0.55\textwidth]{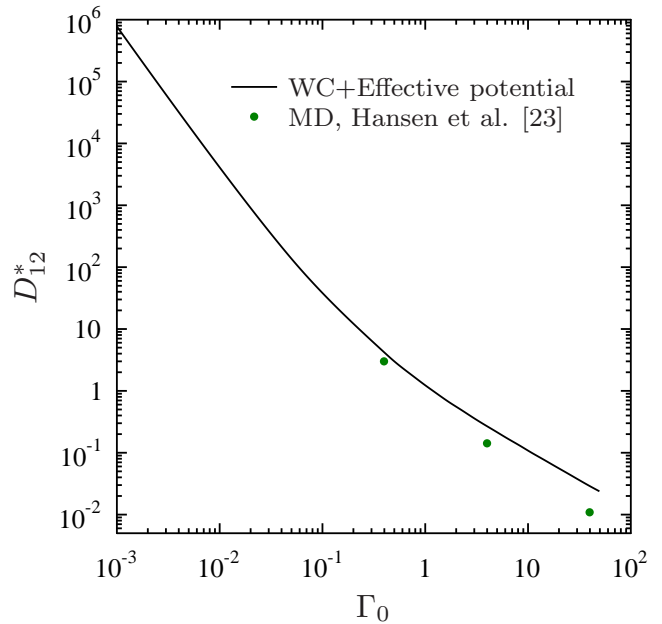}
\caption{(Color online). Interdiffusion coefficient $D_{12}^*$ for
$^{1}$H -- $^{4}$He mixture ($x_1 = 0.5$) and its comparison with MD
data of  Ref.\ \cite{HJM85}. Weak, strong
and intermediate coupling regions are distinguishable (cf. Fig.\ \ref{fig:Lambda}). Exact values of $D_{12}^*$
are given in Table \ref{Tab:Compar}.}
\label{fig:DiffCoef}
\end{figure}

Unfortunately, there is not very much data available to compare our
interdiffusion coefficients with. As seen from Fig.\
\ref{fig:DiffCoef} and Table \ref{Tab:Compar}, the diffusion
coefficients $D_{12}^*$ obtained via the effective potential are systematically
larger than the MD results $D_{12}^{*\mathrm{MD}}$ of \citet{HJM85}, and the difference
increases with increasing $\Gamma_0$. This is exactly the same behavior
as in the original work of Baalrud and Daligault \cite{BD13} (their
Fig.\ 2) who proposed the effective potential method. We have also
compared our data to MD data of Refs.\ \cite{RNZ95,BP87} and
obtained similar results. This seems to be the consequence
of the approximate nature of the effective potential method itself.
Since MD data are  obtained from first principles, they should have been
considered as superior to HNC ones. Nevertheless, the disagreement
between the MD and HNC results appears at strong Coulomb coupling where
quantum effects in ion motion become important. Unfortunately, the quantum
effects are  included neither in the MD nor in the HNC schemes we refer to.
In this situation,
we see no way to check our results with really exact solutions. Therefore, we propose
to use the HNC results, which can be obtained quickly. We do not expect that
the exact solution, if available, would lead to very different diffusion of ions
in liquid BIMs.

Using our (effective potential) $D_{12}$, we have also tried to derive an approximate relation similar to \eqref{e:D12=D11+D22}. Our best attempt gives
$D_{12}\approx D_{12}^\mathrm{S}$, with
\begin{align}
   D_{12}^\mathrm{S} (n,T) \approx x_2 D_1(\widetilde{n}_1,T) + x_1 D_2(\widetilde{n}_2,T),
\label{e:D12=D1+D2}
\end{align}
where $D_1$ and $D_2$ are self-diffusion coefficients in ``equivalent'' OCPs and
\begin{equation}
   \widetilde{n}_j = \frac{\overline{Z^2}}{Z_j^2}~n.
\label{e:nOCP}
\end{equation}
Such a choice of $\widetilde{n}_j$ forces the Debye screening length
in ``equivalent'' OCPs to be the same as in the BIM. This resembles
the linear mixing rule (see, e.g., Ref.\ \cite{HTV77}), where
``equivalent'' OCPs are taken in such a way that they retain the
same electron number density as in a BIM [i.e. $ \widetilde{n}_j =
({\overline{Z}}/{Z_j})n$]. Equation \eqref{e:D12=D1+D2}
was initially obtained semiempirically for weakly coupled plasma, but is
not greatly violated in the strong coupling regime, despite the fact that the concept
of the Debye ion screening length does not apply to strongly coupled
plasma. Examples of $D_{12}^\mathrm{S}$ are presented in Table
\ref{Tab:Compar}.


\section{Conclusions}
\label{s:Conclusions}

We have considered interdiffusion coefficients $D_{12}$ of ions (of two species, 1 and 2) in BIMs under the assumption that the ions consititute either a Boltzmann gas or Coulomb liquid, and the electrons form nearly a uniform background. The problem has been studied for a long time in a number of publications (Sec.\ \ref{s:introduct}), but a unified practical procedure of calculating many diffusion coefficients important for applications has been absent. The main obstacle consisted in substantial computational difficulties
of calculating $D_{12}$ by rigorous methods like MD in the regime of strong Coulomb coupling.

We have used the method of effective inter-ion potential suggested recently by Baalrud and Daligault \cite{BD13}. They proposed to determine the effective potential by a reasonably simple HNC scheme and use this potential to evaluate the diffusion coefficient by the standard Chapman-Enskog method. The latter method is known to be strictly valid for rarefied, weakly coupled plasmas, whereas Baalrud and Daligault
suggested to apply it in both regimes (gas and liquid). They proved that the method is reasonably accurate for calculating the selfdiffusion coefficient of ions in OCP. We have extended their consideration to BIMs and show that the method remains sufficiently accurate for calculating interdiffusion coefficients in BIMs.
The combination of two well-elaborated schemes (the HNC scheme for finding the effective potential and the Chapman-Enskog scheme for evaluating kinetic coefficients) makes this method feasible for determining many interdiffusion coefficients of practical importance in BIMs over wide ranges of temperatures and densities.

To demonstrate the efficiency of this method we have calculated $D_{12}$ for five BIMs
($^1$H--$^4$He, $^1$H--$^{12}$C, $^4$He--$^{12}$C, $^{12}$C--$^{16}$O, $^{16}$O--$^{79}$Se). In analogy with the results of Ref.\ \cite{Khrapak13}, the diffusion coefficients $D_{12}$ have been expressed (\ref{e:Lambda-eff-def}) through a generalized Coulomb logarithm $\Lambda_\mathrm{eff}$. We have approximated all calculated values of
$\Lambda_\mathrm{eff}$ by a unified fit formula
(\ref{e:Lambda-fit}) which contains five fit parameters for each BIM (listed in Table
\ref{Tab:FitParam}). In this way we have obtained a unified description of the interdiffusion coefficients for these BIMs. We may easily consider other BIMs if necessary.

Let us stress once more that in the strongly coupled plasma the
employed effective potential approach \cite{BD13} is phenomenological. We expect that our results can be less accurate in this limit than in the limits of
weak and intermediate Coulomb couplings. However, when the temperature decreases to the melting temperature $T_m$, quantum effects in ion motion can become
important for various properties of the matter (e.g., Ref.\ \cite{HPY07}). In particular, they can affect diffusion, and the effect has not been studied at all, to the best of our knowledge. In this situation (the quantum effects
are neglected anyway) our approach seems reasonable (although the incorporation of quantum effects would be desirable).

Although we have not focused on self-diffusion coefficients in BIMs, we remark
that they are most probably calculated by the effective potential method less accurately than the self-diffusion coefficients in OCP
\cite{BD13}. The nature of this phenomenon is not entirely clear. It
be may because the calculation of self-diffusion coefficients $D_{\mathit{ii}}$ for one component in a BIM requires not only $\Phi_{\mathit{ii}}$, but also
$\Phi_{\mathit{ij}}$, whereas, according to Sec.\ \ref{s:Discussion}, the computation of the interdiffusion coefficient $D_{\mathit{ij}}$  primarily requires only $\Phi_{\mathit{ij}}$.  This problem remains to be solved along the basic problem of why the effective potential is reasonably successful in the regime of strong coupling.

Our results (combined with those of Ref.\ \cite{BY13}) can be used to study various diffusion processes of ions in the crust of neutron stars and in the cores of white dwarfs (e.g. Refs.\
\cite{Potekhin_etal97,CB03,CB04, CB10,IM85,Isern_etal91,BH01,DB02,Althaus_etal10,Garcia_etal10}) as well as in dense Coulomb plasmas of giant and supergiant stars and giant planets.
Such diffusion processes can affect thermodynamics and kinetics of dense matter, thermal and chemical evolution of these stars, and their vibrational properties
(seismology). The diffusion properties of Coulomb plasmas are also important
for dusty plasmas, inertial confinement fusion, etc. (Sec.\ \ref{s:introduct}).

Numerically, our diffusion coefficients are in reasonable agreement with those obtained by other authors and with different techniques (Sec.\ \ref{s:introduct}). The main advantage of our results is in simplicity, uniformity, and convenient approximate expressions. Another important advantage is that the effective potential method can be easily generalized for calculating other kinetic properties of strongly coupled Coulomb plasmas, for instance, the diffusion and thermal diffusion coefficients in multicomponent ion mixtures which are needed for applications but which are almost not considered in the literature. However, we should warn the reader once more
that the method of an effective potential at strong Coulomb coupling is phenomenological
in its essence. It would be important to justify this method and understand the
conditions at which it is most accurate. It would be even more important to
study diffusion in strongly coupled Coulomb plasmas taking into account quantum
effects in ion motions. However, all these difficult issues seem to be beyond the scope
of the present investigation.

Although we have a considered rigid (almost incompressible) electron background, the results can be easily generalized to the case of compressible background produced by electrons of any degeneracy and relativity.

\begin{acknowledgements}
The authors are grateful for the partial support by the State
Program ``Leading Scientific Schools of Russian Federation'' (grant NSh 294.2014.2). The
work of MB has also been partly supported by the Dynasty Foundation,
and the work of DY by Russian Foundation for Basic Research (grants Nos. 14-02-00868-a and
13-02-12017-ofi-M) and by ``NewCompStar'', COST Action MP1304.
\end{acknowledgements}

\end{document}